\documentclass[prl,aps,preprint,tightenlines,showpacs,superscriptaddress]{revtex4}  
\usepackage{epsf}			
\newbox\mybox
\newcommand\fverb{\setbox\mybox=\hbox\bgroup\verb}
\newcommand\fverbdo{\egroup\medskip\noindent\fbox{\unhbox\mybox}\ }
\newcommand\fverbit{\egroup\item[\fbox{\unhbox\mybox}]}

\def\TMG{{\it{\tau\rightarrow\mu\gamma}}}
\def\TEG{{\it{\tau\rightarrow e\gamma}}}
\def\T2{\tau\tau}
\def\Minv{M_{\rm inv}}
\def\DE{\Delta E}

\def\EG{\it{e\gamma}}
\def\notE{\it{e\!\!\! /}}
\def\EEG{ee\gamma}
\def\TTG{\tau\tau\gamma}

\begin{document}
\preprint{\tighten\vbox{\hbox{\hfil Belle-preprint 2005-2}
                        \hbox{\hfil KEK Preprint 2004-93}
                        \hbox{\hfil Intended for {\it PLB}}
                        \hbox{\hfil Author: K.Hayasaka, T. Ohshima}
}}

\title{\vspace*{1cm}Search for {\boldmath $\TEG$} decay at Belle}

\date{\today}

\begin{abstract}
We have searched for the lepton-flavor-violating decay $\TEG$ 
using a data sample of 86.7 $\rm{fb^{-1}}$
 collected  with the Belle 
detector at the KEKB asymmetric ${e^+e^-}$ collider. 
No evidence for a signal is obtained,
and we set an upper limit for the branching fraction 
$\rm{{\cal B}(\TEG) < 3.9\times 10^{-7}}$ at 
the 90\% C.L. 
\end{abstract}


\affiliation{Budker Institute of Nuclear Physics, Novosibirsk, Russia}
\affiliation{Chiba University, Chiba, Japan}
\affiliation{Chonnam National University, Kwangju, South Korea}
\affiliation{University of Cincinnati, Cincinnati, OH, USA}
\affiliation{University of Hawaii, Honolulu, HI, USA}
\affiliation{High Energy Accelerator Research Organization (KEK), Tsukuba, Japan}
\affiliation{Hiroshima Institute of Technology, Hiroshima, Japan}
\affiliation{Institute of High Energy Physics, Chinese Academy of Sciences, Beijing, PR China}
\affiliation{Institute of High Energy Physics, Vienna, Austria}
\affiliation{Institute for Theoretical and Experimental Physics, Moscow, Russia}
\affiliation{J. Stefan Institute, Ljubljana, Slovenia}
\affiliation{Kanagawa University, Yokohama, Japan}
\affiliation{Korea University, Seoul, South Korea}
\affiliation{Kyungpook National University, Taegu, South Korea}
\affiliation{Swiss Federal Institute of Technology of Lausanne, EPFL, Lausanne, Switzerland}
\affiliation{University of Ljubljana, Ljubljana, Slovenia}
\affiliation{University of Maribor, Maribor, Slovenia}
\affiliation{University of Melbourne, Victoria, Australia}
\affiliation{Nagoya University, Nagoya, Japan}
\affiliation{Nara Women's University, Nara, Japan}
\affiliation{National Central University, Chung-li, Taiwan}
\affiliation{National United University, Miao Li, Taiwan}
\affiliation{Department of Physics, National Taiwan University, Taipei, Taiwan}
\affiliation{H. Niewodniczanski Institute of Nuclear Physics, Krakow, Poland}
\affiliation{Nihon Dental College, Niigata, Japan}
\affiliation{Niigata University, Niigata, Japan}
\affiliation{Osaka City University, Osaka, Japan}
\affiliation{Osaka University, Osaka, Japan}
\affiliation{Panjab University, Chandigarh, India}
\affiliation{Peking University, Beijing, PR China}
\affiliation{Princeton University, Princeton, NJ, USA}
\affiliation{University of Science and Technology of China, Hefei, PR China}
\affiliation{Seoul National University, Seoul, South Korea}
\affiliation{Sungkyunkwan University, Suwon, South Korea}
\affiliation{University of Sydney, Sydney, NSW, Australia}
\affiliation{Tata Institute of Fundamental Research, Bombay, India}
\affiliation{Toho University, Funabashi, Japan}
\affiliation{Tohoku Gakuin University, Tagajo, Japan}
\affiliation{Tohoku University, Sendai, Japan}
\affiliation{Department of Physics, University of Tokyo, Tokyo, Japan}
\affiliation{Tokyo Institute of Technology, Tokyo, Japan}
\affiliation{Tokyo Metropolitan University, Tokyo, Japan}
\affiliation{Tokyo University of Agriculture and Technology, Tokyo, Japan}
\affiliation{University of Tsukuba, Tsukuba, Japan}
\affiliation{Virginia Polytechnic Institute and State University, Blacksburg, VA, USA}
\affiliation{Yonsei University, Seoul, South Korea}
\author{K.~Hayasaka}\affiliation{Nagoya University, Nagoya, Japan}
\author{K.~Abe}\affiliation{High Energy Accelerator Research Organization (KEK), Tsukuba, Japan}
\author{K.~Abe}\affiliation{Tohoku Gakuin University, Tagajo, Japan}
\author{H.~Aihara}\affiliation{Department of Physics, University of Tokyo, Tokyo, Japan}
\author{Y.~Asano}\affiliation{University of Tsukuba, Tsukuba, Japan}
\author{T.~Aushev}\affiliation{Institute for Theoretical and Experimental Physics, Moscow, Russia}
\author{S.~Bahinipati}\affiliation{University of Cincinnati, Cincinnati, OH, USA}
\author{A.~M.~Bakich}\affiliation{University of Sydney, Sydney, NSW, Australia}
\author{I.~Bedny}\affiliation{Budker Institute of Nuclear Physics, Novosibirsk, Russia}
\author{U.~Bitenc}\affiliation{J. Stefan Institute, Ljubljana, Slovenia}
% ...
\collaboration{Belle Collaboration}

\maketitle

\section{Introduction}
Lepton-flavor-violating (LFV) processes are  good probes of 
physics beyond the Standard Model (SM). 
For instance, in some supersymmetric models, off-diagonal 
components of the left-handed slepton mass matrix, $m_{\tilde{L}}$, 
could radiatively induce LFV such as in 
$\tau\rightarrow\mu (e)\gamma$ and $\mu\rightarrow e\gamma$ 
decays \cite{Barbieri:1994pv,hisano:1995}.
In general, the branching fraction ${\cal{B}}(\TMG)$ is 
expected to be larger than ${\cal{B}}(\TEG)$, since the mixing between 
the third and second families is typically assumed to be stronger 
than that between the third and first families. However, 
if the first and third families couple more strongly,
 for instance, due to 
an inverted hierarchy of slepton masses,
then ${\cal{B}}(\TEG)$ could exceed ${\cal{B}}(\TMG)$
and might be detectable~\cite{ellis}.
Values of ${\cal B}(\TEG)$ which can exceed that for 
${\cal B}(\TMG)$ are also
predicted in the models with heavy Dirac neutrinos~\cite{Gonza:1992,Ila:2000}.
Thus, a study of both $\TEG$ and $\TMG$ decays is essential not 
only to search for new physics but also to 
further examine lepton flavor structure.

The decay $\TEG$ has been searched for, along with $\TMG$, by 
MARK II~\cite{mark2}, Crystal Ball~\cite{cryst}, ARGUS~\cite{args}, 
DELPHI~\cite{delf}, and CLEO~\cite{cleo1}, 
among which CLEO has set the most sensitive upper limit of 
${\cal B}(\TEG) < 2.7 \times 10^{-6}$ at 90\% C.L.

Recently the Belle collaboration performed a search
for the LFV decay $\tau \rightarrow \mu\gamma$~\cite{MuG}. 
Here we present
a new search for the decay $\tau \to e \gamma$ based on data samples of
77.7 fb$^{-1}$ and 9.0 fb$^{-1}$,
collected at the $\Upsilon$(4S)
resonance and in the continuum 60~MeV below the resonance,
respectively,
equivalent in total to $77.3\times10^6$ $\tau^+\tau^-$ pairs.
The data were collected with the Belle detector 
at the KEKB asymmetric $e^+e^-$ collider~\cite{kekb}.
A description of the detector can be found in Ref.~\cite{Belle}.

\section{Data Selection}

 We search for events containing exactly two oppositely-charged
tracks and at least one photon. The events should be consistent
with a $\tau^+\tau^-$ event in which one $\tau$ (signal side) decays 
to $e \gamma$
and the other (tag side)   decays to a non-electron
charged particle (denoted hereafter as $\notE$), neutrino(s) and
any number of photons.
 
The selection criteria are determined by studying  Monte Carlo (MC)
simulations for signal $\tau$-pair decay and
background (BG) events, such as generic  $\tau$-pair decay 
($\tau^+\tau^-$), $q\bar{q}$ continuum, $B\bar{B}$, Bhabha, 
$\mu^+\mu^-$, and two-photon events~\cite{MuG}. 
The KORALB/TAUOLA~\cite{KORALB} and QQ~\cite{QQ} generators are used
for event generation, and the Belle detector response is simulated by 
a GEANT3~\cite{GEANT} based program. The two-body decay $\TEG$ is 
initially assumed to have a uniform angular distribution in the 
$\tau$ lepton's rest system.

The selection criteria are similar to those used in the $\TMG$ 
search~\cite{MuG}. 
Kinematic variables  with a CM superscript  are calculated 
in the center-of-mass frame; all other variables  are calculated 
in the laboratory frame.
Before electron identification,
 all the charged tracks are assumed to be pions.
Each track is required to have momentum $p^{\rm CM} <$ 4.5 GeV/$c$ 
and momentum transverse to the $e^+$ beam $p_t >$ 0.1 GeV/$c$, 
the former requirement being imposed
to avoid Bhabha and $\mu^+\mu^-$ contamination. 
We require that the energy $E_{\gamma}$ of each photon
exceed 0.1 GeV.
In addition, we also require the total energy measured in
the CsI(Tl) electromagnetic calorimeter (ECL),
$E_{\rm ECL}$, to be less than 9 GeV in order to
suppress background from Bhabha events.

The tracks and photons must be detected within the
detector's fiducial volume $ -0.866 < \cos\theta < 0.956$,
but outside the barrel--endcap gaps
defined by
 $ 0.829 < \cos\theta < 0.880$ and
 $ -0.716 < \cos\theta < -0.602$.
Here, $\theta$ is the polar angle with respect to
the direction opposite to the $e^+$ beam.
Identification of electrons is performed using
an electron likelihood ratio, 
${\cal L}_e$, which is based on the $dE/dx$ information
from the central drift chamber (CDC), the ratio of the energy
deposited in the ECL to the
momentum measured by both the CDC
and the silicon vertex detector (SVD), 
the shower shape in the ECL, 
the hit information from the aerogel Cherenkov counter,
and time-of-flight measurements~\cite{Hanagaki:2001fz}.

\begin{figure*}
\begin{center}
\epsfxsize=16.0cm \epsfbox{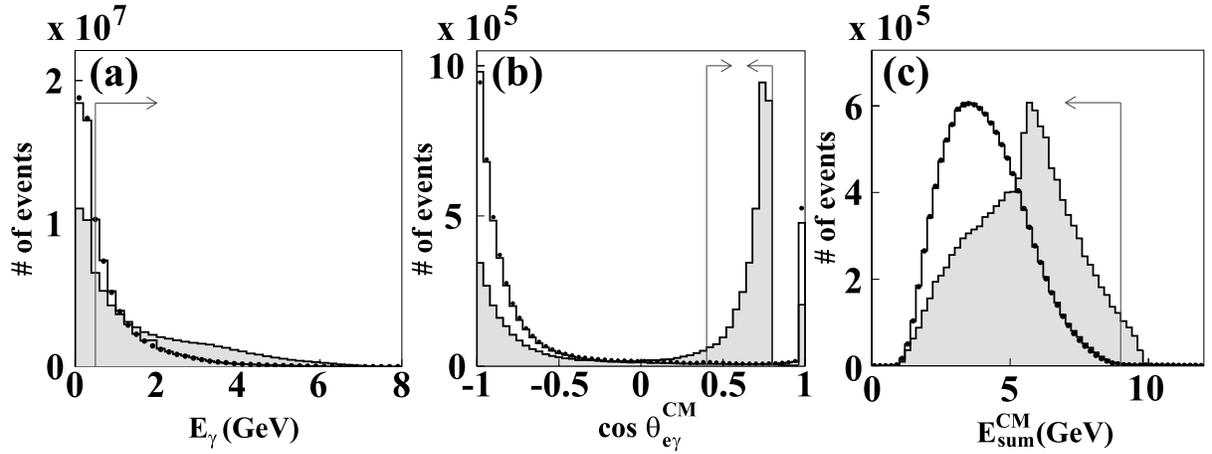} 
\end{center}
\vspace*{-5mm}
\caption{ (a) Energy distribution of the signal candidate photon.
(b) $\cos\theta_{e\gamma}^{\rm CM}$ distribution.
(c) $E_{\rm sum}^{\rm CM}$ distributions.
The open histogram is the sum of backgrounds from
 generic $\tau^+\tau^-$, $q\bar{q}$ (uds) continuum, 
Bhabha, $\mu^+\mu^-$ and two-photon processes
evaluated from MC simulation.
Dots indicate the data distribution, and the shaded
histogram is the signal MC distribution.  
Electron identification requirements were applied for these figures.
}
 \label{Fig-1d}
\end{figure*}

The electron track that forms a $\TEG$ candidate (hereafter denoted 
as $(\EG)$) is required to have an $e$ likelihood ratio
${\cal L}_{\rm e} >$ 0.90 and a momentum
$p >$ 1.0 GeV/$c$. This requirement has an 
 efficiency of $(93\pm3)$\%
in the barrel and forward detector and $(76\pm7)$\%
in the backward detector
because of the additional material.
On the tag side, the $\notE$ track is required to have 
${\cal L}_{\rm e} <$ 0.1. The fraction  $\eta$ of electrons with 
${\cal L}_{\rm e} <$ 0.1 is measured to
be $(4\pm3)$\% in the barrel and forward detector and $(13\pm5)$\%
in the backward detector for $p > 1.0$ GeV/$c$.

The photon that forms an $(\EG)$ candidate is required to have 
$E_{\gamma} >$ 0.5 GeV 
in order to reduce spurious combinations of a low-energy $\gamma$ 
with an electron, see Fig.~\ref{Fig-1d}(a).

A requirement on the cosine of the opening angle 
between the $e$ and $\gamma$ of 
the $(\EG)$ candidate, 0.4 $< \cos\theta^{\rm CM}_{\EG} <$ 0.8,
is particularly  powerful in rejecting the generic 
$\tau^+\tau^-$ BG events (see Fig.~\ref{Fig-1d}(b)).
The events in  Fig.~\ref{Fig-1d}(b)
that peak at $\cos\theta^{\rm CM}_{e\gamma}\sim 1$,
arise from electrons that radiate a photon when they interact
in the SVD or in materials around it.
The requirement $E^{\rm CM}_{\rm sum} <$ 9.0 GeV   is imposed to reject 
Bhabha and $\mu^+\mu^-$ production, where $E^{\rm CM}_{\rm sum}$
is defined as the sum of the energies of the two charged tracks and 
the photon composing the $(\EG)$, see Fig.~\ref{Fig-1d}(c).
The opening angle between the two tracks in the laboratory frame
is required to be greater than 90$^{\circ}$.

We define ${\vec p}_{\rm miss}$ as the residual momentum vector
calculated by subtracting the vector sum of all visible momenta 
(of both tracks and photons) from the vector sum of the beam momenta.
Constraints on the momentum and cosine of the polar angle of the missing
particle are imposed: $p_{\rm miss} >$ 0.4 GeV/$c$ and
$-0.866 < \cos{\theta}_{\rm miss} < 0.956$. 
To remove $\tau^+\tau^-$ BG events, 
we apply a requirement on the opening angle 
between the tagging track and the missing particle of
 0.4 $< \cos\theta^{\rm CM}_{{\rm miss}-\notE} <$ 0.99.

\begin{figure*}[t]
\begin{center}
\epsfxsize=15cm \epsfbox{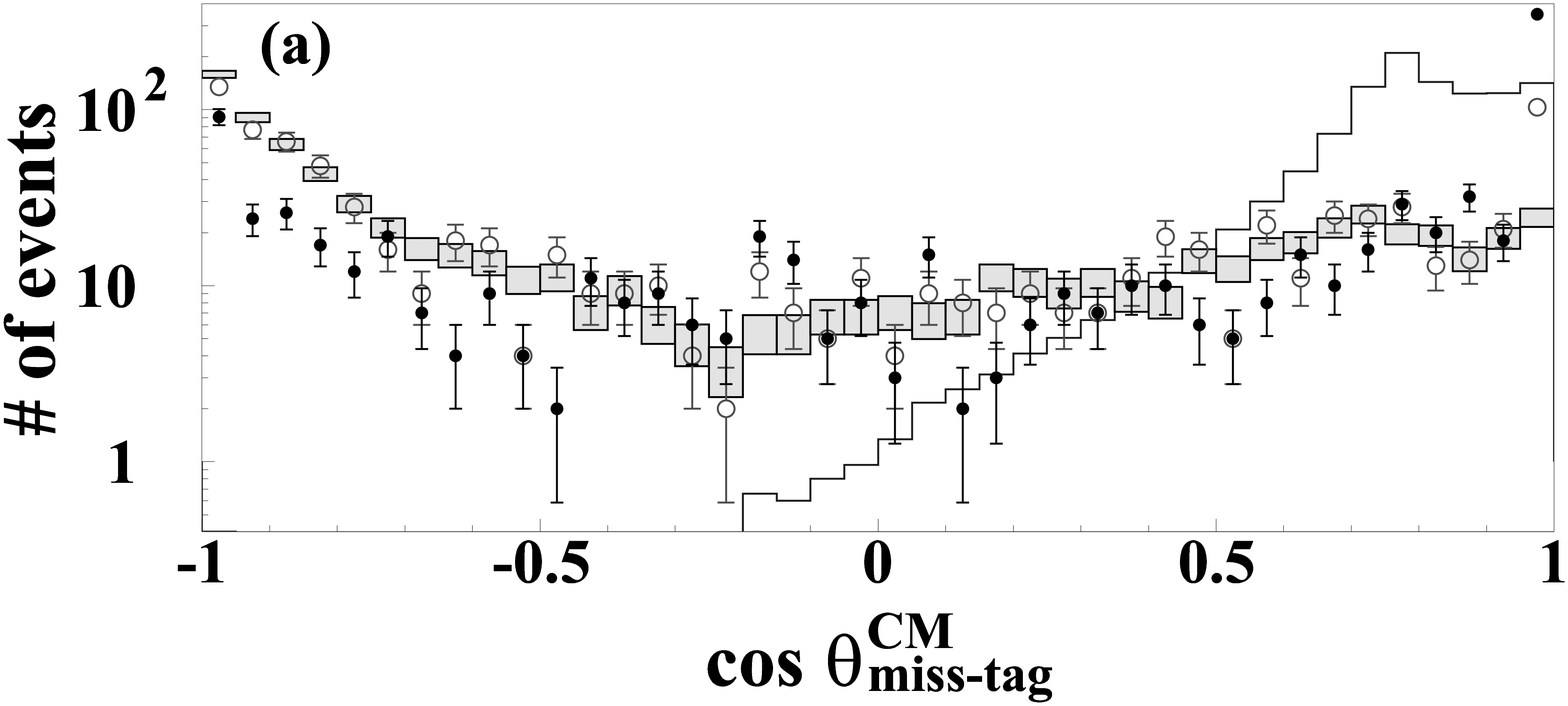}\\
\epsfxsize=12cm \epsfbox{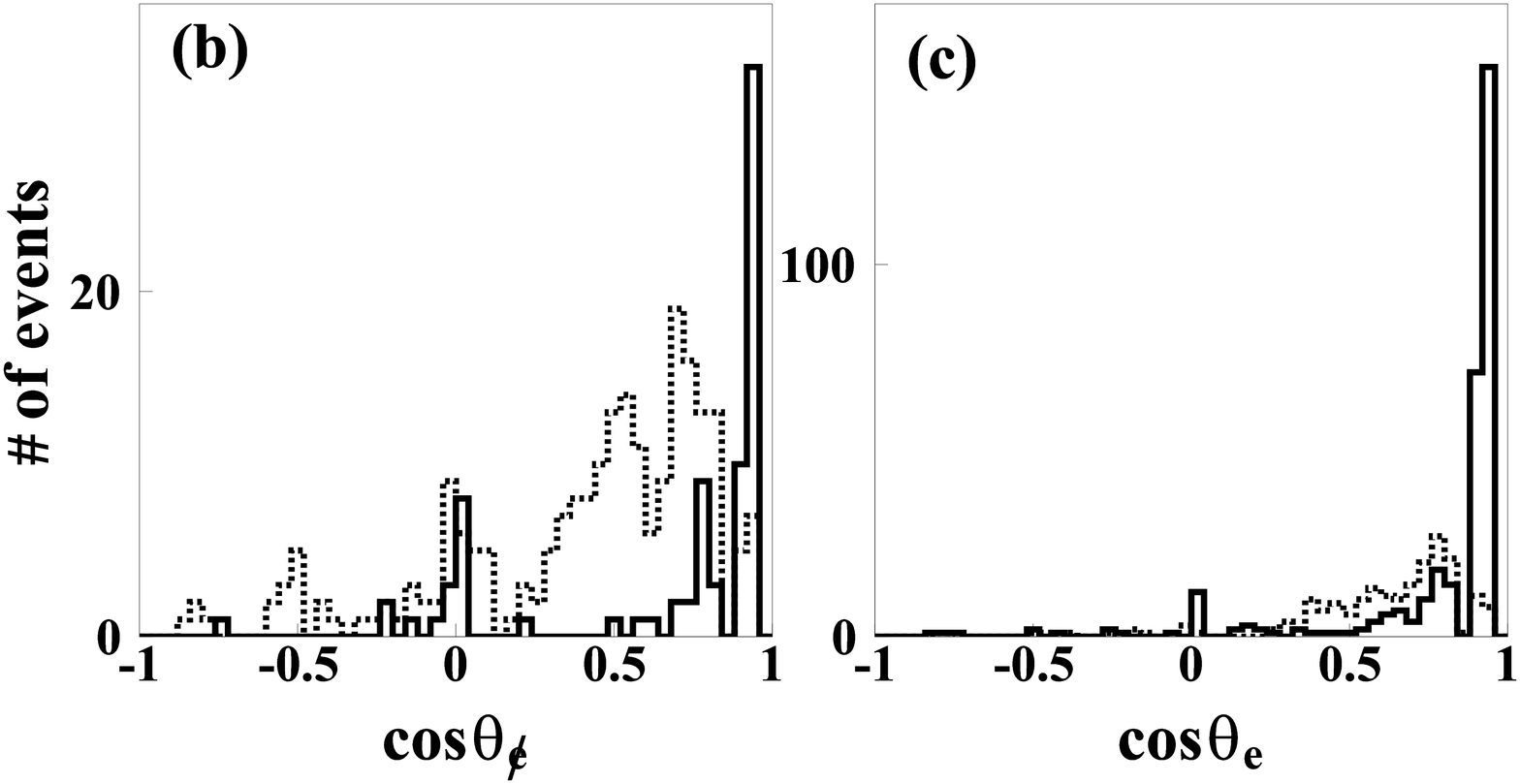}
\end{center}

\caption{
(a) $\cos\theta^{\rm {CM}}_{{\rm miss-tag}}$ distribution.
For $\notE$-tagged events,
the distributions of signal MC (histogram), generic $\tau^+\tau^-$ MC (boxes), 
 and $\notE(\EG)$data sample (open circles) are shown.
For $e$-tagged events,
the distribution for $e(\EG)$ data (closed circles) are also plotted. 
All requirements except the one for $\cos \theta_{\rm miss-tag}^{\rm CM}$ are
applied.
(b) $\cos\theta_{\notE}$ and
  (c) $\cos\theta_{e}$ distributions.
  These are polar angle distributions of the tag side track for
  $\notE(\EG)$ and ${\it{e}}(\EG)$ data, respectively,
  where the solid histogram is for the events with 
  $\cos{\theta}^{\rm {CM}}_{{\rm miss}-{\rm tag}} > 0.99$ and the dotted 
  histogram is for 
$0.4<\cos{\theta}^{\rm {CM}}_{{\rm miss}-{\rm tag}} < 0.99$
(tagged by $e$ or $\notE$).}
\label{Fig-2}
\end{figure*}

The upper bound on $\cos{\theta}^{\rm CM}_{{\rm miss}-\notE}$ is 
introduced to reject radiative Bhabha events in which one of the 
electrons forms an $(\EG)$ candidate with  a radiated photon and 
the electron on the tag side is misidentified as the $\notE$ 
due to the electron identification inefficiency. 
By analyzing a Bhabha data sample, a large portion of such events is 
found to have a very small opening angle, 
$\cos{\theta}^{\rm CM}_{{\rm miss}-\notE}\simeq$ 1, and 
a polar angle peaking strongly forward, $\cos\theta_{\notE} >$ 0.8. 
Figure~\ref{Fig-2}(a) shows the 
$\cos{\theta}^{\rm CM}_{{\rm miss}-{\rm tag}}$ 
distributions with tag given by ${\notE}$ or $e$ for $\notE(\EG)$ or 
${\it e}(\EG)$ modes, respectively, in the actual Bhabha data samples, 
and the signal and generic $\tau^+\tau^-$ MC data. 
Figures~\ref{Fig-2} (b) and (c) present the 
$\cos{\theta}_{\rm{tag}}$ distribution 
for $\notE(\EG)$ and ${\it{e}}(\EG)$ Bhabha data samples,
respectively.
The requirement, $\cos{\theta}^{\rm CM}_{{\rm miss}-{\rm tag}} <$ 0.99,
 reduces 
$\notE(\EG)$ and ${\it{e}}(\EG)$   candidates that originate from
radiative Bhabhas by 73\% and 45\%, respectively, while
only slightly affecting the signal (97\%) and generic 
$\tau^+\tau^-$ (99\%) events.  

\begin{figure*}
\epsfxsize=14cm \epsfbox{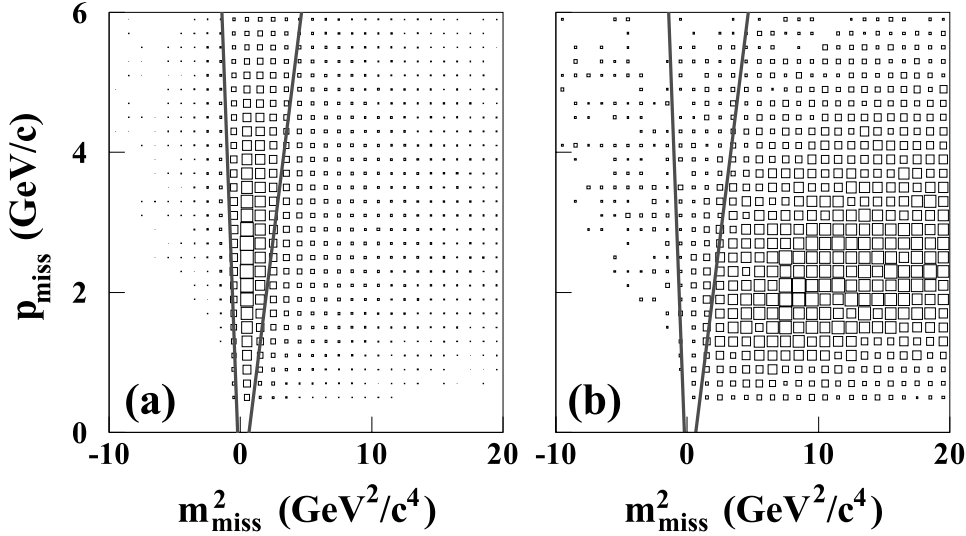}

\caption{Event distribution in the 
$m^2_{\rm miss}- p_{\rm miss}$ plane for (a) signal 
  and (b) generic $\tau^+\tau^-$ MCs. 
  The events within the two lines are accepted in the analysis. }
\label{Fig-3}
\end{figure*}

Finally, a condition is imposed on the relation between 
$p_{\rm miss}$ and the mass-squared of a missing particle, 
$m^2_{\rm miss}$. The latter is defined as 
$E^2_{\rm miss} - p^2_{\rm miss}$, where $E_{\rm miss}$ is 11.5 GeV
(the sum of the beam energies) minus the sum of all visible energy 
and is calculated assuming the electron (pion) mass
for the charged track on the signal (tag) side. We require
$p_{\rm miss} > -5 \ (c^3/{\rm GeV}) \times m^2_{\rm miss}-1 \ ({\rm GeV}/c)$
 and 
$p_{\rm miss} > 1.5 \ (c^3/{\rm GeV}) \times m^2_{\rm miss}-1 \ ({\rm GeV}/c)$, 
where $p_{\rm miss}$ is
in GeV/$c$
and $m_{\rm miss}$ is in GeV/$c^2$  (see Fig.~\ref{Fig-3}).  
With this cut, 98\% of the generic $\tau^+\tau^-$ and 97\% of 
the $e^+e^-\gamma$ backgrounds are removed, 
while 69\% of the signal events remain. 
In addition, most of the remaining $B\bar{B}$, continuum, and 
two-photon events are rejected by this requirement. 

After these selection requirements, {224} events remain
in the data, about 3 times fewer than in the $\TMG$ case.
Since the inefficiency of electron identification is much 
smaller than that of the muon, 
the Bhabha BG is strongly suppressed in spite of
its much larger cross-section than that of $e^+e^-\rightarrow
\mu^+\mu^-\gamma$.
The $\TEG$ detection efficiency is evaluated by MC to be 7.29\%,  
about 40\% smaller than that of $\TMG$, mostly because 
of the  $E_{\rm ECL}$ requirement.

True signal events will have an invariant mass ($\Minv$)
close to the $\tau$ lepton mass and an energy close to the beam energy
in the CM frame, i.e., 
$\DE= E^{\rm CM}_{\it e\gamma}-E^{\rm CM}_{\rm beam} \simeq 0$.
When deciding on our selection criteria, we excluded the signal region
1.68 GeV/$c^2 < \Minv <$ 1.85 GeV/$c^2$ so as not to bias our choice 
of criteria (a ``blind'' analysis). Only after all requirements 
were finalized
and the number of BG events estimated did we include this
region and count the number of signal events.

\begin{figure*}
\begin{center}
\epsfxsize=9cm \epsfbox{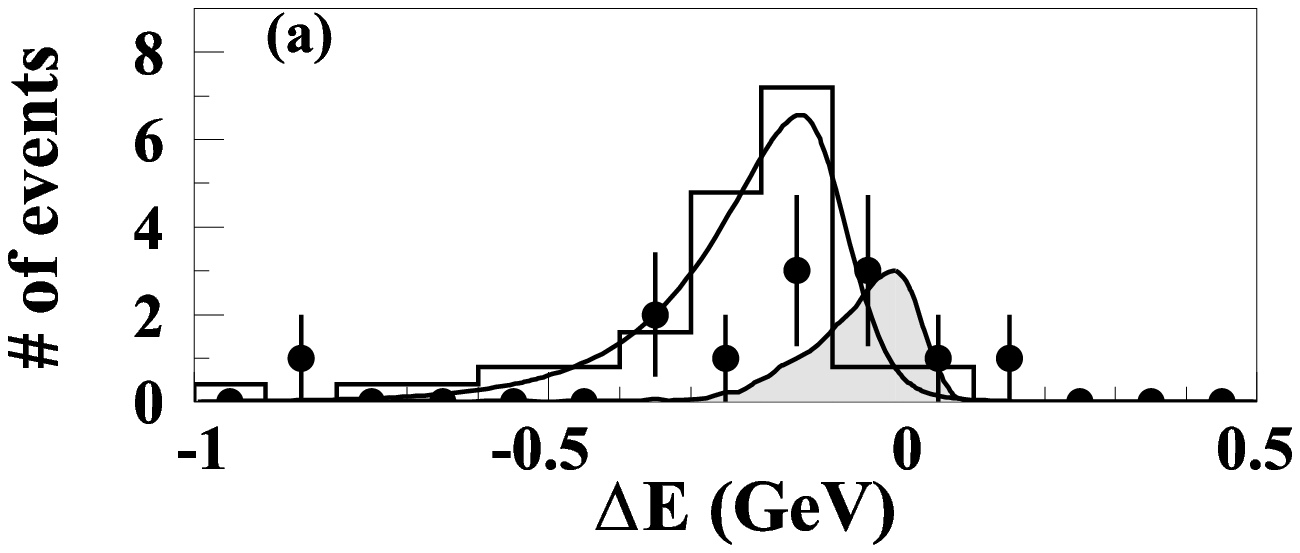}
\epsfxsize=8.7cm \epsfbox{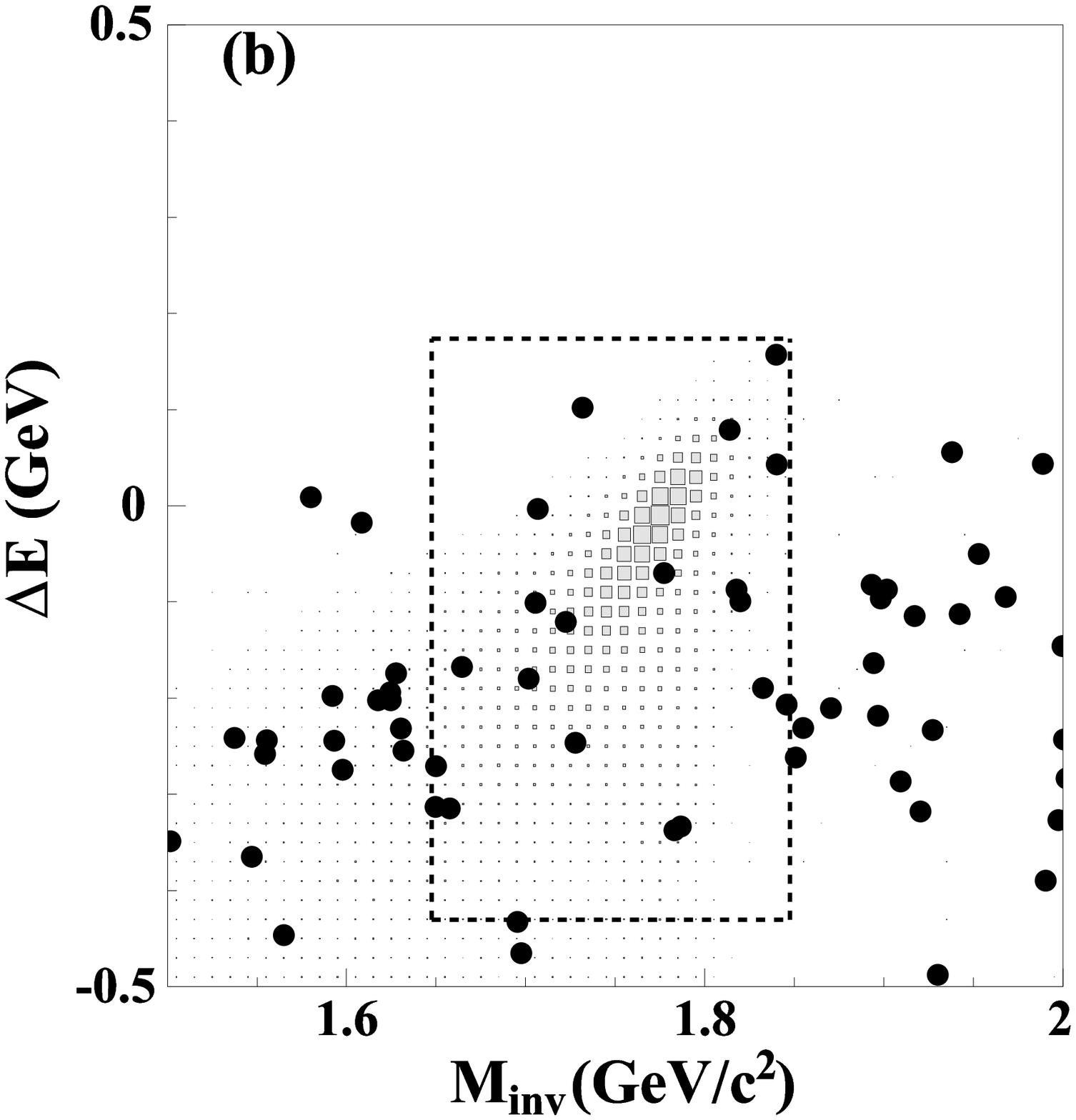}
\makebox[5mm]{}
\vspace*{-0.5cm}
\end{center}
\caption{
(a) $\Delta{E}$ distributions for the data (dots)
and the expected BG (curve and open histogram) in the blinded region.
The distribution for signal MC is the shaded curve.
See the text for more detail.
(b) $\Minv$ {\it vs.} $\Delta{E}$ distributions for the data (dots)
and signal MC (shaded boxes).
The $\pm5\sigma$ region is indicated by the dashed rectangle.
}
\label{Fig-4}
\end{figure*}

\section{Results}
\subsection{Background Evaluation}
To analyze the BG distributions, we define a region, named ``grand signal region'':
$-0.5$ GeV$ < \DE < 0.5~{\rm GeV}$ and 
$1.5~{\rm GeV}/c^2 < \Minv < 2.0~{\rm GeV}/c^2$, containing 90\% 
of signal MC events passing all previous requirements.  

The resolution in $\DE$ and $\Minv$ is evaluated by MC: 
an asymmetric Gaussian reproduces the dominant part of the signal MC 
distribution with 
$\sigma^{\rm{low/high}}_{\DE}$ = $(84.8 \pm 1.2)/(36.0 \pm 0.9)$ MeV
and 
$\sigma^{\rm{low/high}}_{\Minv}$ = $(25.7 \pm 0.3)/(14.3 \pm 0.2)$ 
MeV/$c^2$, where $\sigma^{\rm{low/high}}$ means the standard deviation
at the lower/higher side of the peak. The peak positions are  
$-6.2 \pm 1.0$ MeV and $1776.0 \pm 0.2$ MeV/$c^2$ for $\DE$ and
$\Minv$, respectively.

A dominant source of BG is the process $e^+e^- \to \tau^+\tau^-\gamma$, 
in which the
photon is radiated from the initial state: an $(\EG)$ candidate is
formed by the electron from the $\tau\rightarrow e\nu\overline{\nu}$ 
decay and the initial state radiation photon, while the tag side
 $\tau$ decays generically via a one-prong mode but not to an
electron. From a 174 fb$^{-1}$ sample of MC  $\tau^+\tau^-\gamma$ 
events we find
$N^{\TTG} = {60.8 \pm 5.5}$ events in the ``grand signal region.'' 

The contribution from the process $\notE\EG$
was described above and
is evaluated as $N^{\notE e\gamma} = \kappa\times N^{\EEG}$,
where $\kappa =  \eta /(1-\eta)$. 
From the data $N^{\EEG}$ is found to be ${68.0 \pm 8.2}$ events
and $\kappa$ is estimated to be $0.06 \pm 0.03$ from
both the Bhabha data and MC samples taking into account
 the momentum dependence based on the momentum 
 distribution of the signal MC events.
Thus, we have $N^{\notE\EG} = 4.3 \pm 2.0$ events.

From the MC simulation, no other process is expected to contribute
to the background. 
Therefore, the expected BG  in the ``grand signal region'' is
${65.1 \pm 5.9}$ events.

The $\Minv$ and $\DE$ shapes of
both types of BG events are empirically reproduced 
by a combination of Landau and Gaussian functions. 

For $\tau^+\tau^-\gamma$,
\begin{eqnarray}
N^{\TTG}(\Minv, \DE)
&=&  \left\{ 
\begin{array}{lrlr}
a(\Minv) \times \exp{\left[\displaystyle 
-\left (\displaystyle  \frac{\displaystyle  \alpha}
{\displaystyle \sqrt{2}~\upsilon_{\rm {h}}} 
\right )^2 \right]} 
\vspace*{4mm}\\
\hspace*{2.5 cm}{\rm {for}}~\DE > \DE^{\TTG}_{\rm {peak}}(\Minv),  
\vspace*{3mm}\\
a(\Minv) \times \exp{\left[\displaystyle 
 \displaystyle  \frac12 + \frac{1}{2} 
\left \{
\displaystyle 
  \frac{\displaystyle \alpha}{\displaystyle  \upsilon_{\rm {l}}} -  
\exp \left({\displaystyle \frac{\displaystyle \alpha}
{\displaystyle \upsilon_{\rm {l}}} } \right ) \right \}\right]} 
\vspace*{4mm}\\
\hspace*{2.5 cm} {\rm {for}}~\DE < \DE^{\TTG}_{\rm {peak}}(\Minv),  
\vspace*{3mm}\\
\end{array}
\right.
\end{eqnarray}
and for $e^+e^-\gamma$,
\begin{eqnarray}
N^{\EEG}(\Minv, \DE)
& =  &
b(\Minv) \times \exp{\left [ \displaystyle  \frac12+\frac{1}{2} \left \{
\displaystyle   \frac{\displaystyle \beta}
{\displaystyle \omega_{\rm {h/l}}} -  
\exp\left({\displaystyle \frac{\displaystyle  \beta}{\displaystyle \omega_{\rm {h/l}}} } \right ) \right \}\right]} \nonumber 
\\
\nonumber \\
&&\hspace*{3.5 cm} {\rm {for}}~\DE~ ^{>}_{<}~ 
\DE^{\TTG}_{\rm {peak}}(\Minv).
\end{eqnarray}
Here $\alpha = \DE - \DE_{\rm peak}^{\TTG}(\Minv)$
     and $\beta = \DE - \DE_{\rm peak}^{\EEG}(\Minv)$, where
     $\Delta E_{\rm peak}$ denotes the peak position in terms of 
     $c \times \Minv + d$ for individual BG's.
     The parameters $a, b, c, d, \upsilon_{\rm l/h}$ and 
$\omega_{\rm l/h}$
are determined from
MC for $\tau^{+}\tau^{-}\gamma$ 
and from data for the $e^+e^-\gamma$.

The BG distribution
can then be represented
by the sum of the two BG components above as 
\begin{equation}
  N_{\rm {BG}}(\Minv, \DE) = N^{\TTG}(\Minv, \DE) 
          + \kappa\times N^{\EEG}(\Minv, \DE). 
\label{eq43}
\end{equation}

Figure~\ref{Fig-4}(a) compares the $\DE$ distribution in the  
$1.70$ GeV/$c^2 <\Minv< 1.82$ GeV/$c^2$ ($\pm 3\sigma_{\Minv}$) region 
for BG events expected 
from  Eq.~(\ref{eq43}) (the solid curve) and the events obtained 
by interpolating the data distribution from 
both sidebands, $1.53$ GeV/$c^2 < \Minv < 1.68$ GeV/$c^2$ and 
$1.85$ GeV/$c^2 < \Minv < 2.0$ GeV/$c^2$ (the open histogram). 
Good agreement between them is observed.

\subsection{Upper Limit for {\boldmath ${{\cal B}(\tau\rightarrow{e}\gamma)}$}}
After opening the blinded region, 
we find the $\DE$ and $\Minv$ {\it vs.} $\DE$
distributions that are shown in 
Figs.~\ref{Fig-4} (a) and (b), respectively.
The number of surviving data events in the ``grand signal region'' is 60,
in good agreement with the expected BG contribution of
 $65.1\pm5.9$ events.

In order to extract the number of signal events from the surviving
sample, we apply an unbinned extended maximum likelihood fit with
the likelihood function  defined as 
\begin{equation}
{\cal {L}} = \frac{ e^{-(s+b)}}{ {N!}} \prod_{i=1}^{ {N}}
  (s S_i + b B_i), 
\end{equation}
where $N$ is the number of observed events,
$s$ and $b$ are free parameters representing 
the numbers of signal and BG events
 to be extracted, respectively, and $S_i\equiv S(\Minv^{(i)}, \DE^{(i)})$ and 
$B_i\equiv B(\Minv^{(i)}, \DE^{(i)})$ are the  signal and BG
probability density functions for the $i$-th event. 
The function $B(\Minv, \DE)$ is taken from Eq.~(\ref{eq43})
while $S(\Minv, \DE)$ is 
obtained by generating 10$^6$ signal MC events. 

We apply this fit for $s$ and $b$ to a $\pm 5 \sigma$ region in  
$\Minv$ and $\DE$: 
1.65~GeV/$c^2 <\Minv < $1.85~GeV/$c^2$ and 
$-0.43$~GeV $ < \DE < $0.17~GeV. 
There are a total of  ${20}$ events in 
this region
while $25.7\pm0.3$ events are expected from Eq.~(\ref{eq43}),
and, when $s$ is constrained to be non-negative, 
the fit finds $s = 0$ and $b = 20.0$.

To calculate the upper limit, Monte Carlo samples are generated  
by fixing the expected number of BG events ($\tilde{b}$) to the value 
$b = 20$. 
For every assumed expected number of signal events ($\tilde{s}$), 
10,000 samples are generated, for each of which
the numbers of signal and BG events are determined by
 Poisson statistics with 
means $\tilde{s}$ and $\tilde{b}$, respectively. 
We then assign $\Minv$ and 
$\DE$ values to these events according to their density distributions. 
An unbinned maximum likelihood fit is performed for every sample to 
extract the number of signal events ($s^{\rm {MC}}$). 
The confidence level for an assumed $\tilde{s}$ is defined as 
the fraction of the samples whose $s^{\rm {MC}}$ exceeds $s$. 
This procedure is repeated until we find the value of $\tilde{s}$ 
($\tilde{s}_{90}$) 
that gives a 90\% chance of $s^{\rm {MC}}$ being larger than $s$. 

The resulting upper limit at 90\% C.L. is $\tilde{s}_{90} = 3.75$ events. 
An upper limit on the branching fraction is obtained via the formula: 
\begin{equation}
{\cal{B}}(\TEG) < \frac{\tilde{s}_{90}}{2 \epsilon N_{\T2}}, 
\end{equation}
where $N_{\T2}$ is the total number of $\tau$-pairs produced, 
and $\epsilon$ is the detection efficiency in the $\pm 5 \sigma$ region. 
Inserting the values 
$N_{\T2} = 77.3 \times 10^6$ and $\epsilon = 6.37\%$
gives ${\cal{B}}(\TEG) < 3.8 \times 10^{-7}$.

\subsection{Systematic Uncertainties}
Systematic uncertainties on $\tilde{s}_{90}$ are evaluated 
by varying all 
parameters of the BG probability density function.
The fractions of $N^{\TTG}(\Minv, \DE)$ and $N^{\EEG}(\Minv, \DE)$ 
in Eq.~(\ref{eq43}) are varied by $\pm 20\%$ and 
$\pm 100\%$, respectively, 
about double their estimated uncertainties.
As a result, $\tilde{s}_{90}$ varies by $+0.01$/$-0.00$ and 
$+0.01$/$-0.02$ events, respectively. 
The functional form of the BG spectra is 
scaled by 1.15 or 0.90 times for $N^{\TTG}(\Minv,\DE)$ 
and by 1.3 or 0.6 times for $N^{\EEG}(\Minv, \DE)$, and their centers 
are shifted by $+0.01$/$-0.015$ GeV for $N^{\TTG}(\Minv, \DE)$ and 
by $\pm 0.1$ GeV for $N^{\EEG}(\Minv, \DE)$, all changes corresponding
to the estimated errors of the involved parameters. 
The shift of the central value for the $N^{\TTG}(\Minv, \DE)$ 
spectrum yields the largest effect of $+0.07$/$-0.13$ events,
and the overall systematic uncertainty 
increasing the upper limit of $\tilde{s}_{90}$
is evaluated as $+0.07$ events.
The stability of the result for the fit region is examined
by extending the $\Minv$-$\DE$ region from 
{$\pm 4\sigma$} to $\pm 8\sigma$: 
no appreciable difference in the upper limit is found. 

The systematic uncertainties on the detection sensitivity, 
$2\epsilon N_{\T2}$, arise from the photon reconstruction efficiency (3.0\%),
the selection criteria (2.5\%),
the trigger efficiency (2.0\%), 
the track reconstruction efficiency (2.0\%),
the luminosity (1.4\%), 
and the MC statistics (0.3\%). 
The total uncertainty is obtained by adding all of these components 
in quadrature; the result is 5.0\%. 
The contribution of the largest component,
the photon reconstruction efficiency,
is evaluated from the $e^+e^-\gamma$ data sample.
The uncertainty of the selection criteria is estimated by varying
the required polar angle region of the signal candidate photon.
The trigger efficiency is estimated from the difference
between a $\tau^+\tau^-$ data sample and a generic $\tau^+\tau^-$
MC sample.

These uncertainties are included in the upper limit on ${\cal{B}}(\TEG)$ 
following~\cite{CLEO2}. 
The systematic uncertainty in the efficiency is assumed
to have a Gaussian distribution.

While the angular distribution of the $\TEG$ decay is assumed to be 
uniform in this analysis,
it is sensitive to the LFV interaction 
structure \cite{Kitano}, and spin correlations between 
the $\tau$ leptons on the signal and tag sides 
must be considered. 
To evaluate the maximum possible variation, $V-A$ and $V+A$ 
interactions are assumed; no statistically significant difference 
in the $\Minv$-$\DE$ distribution or in the efficiency is found 
compared to the case of the uniform distribution. Therefore, 
systematic uncertainties due to these effects are neglected
in the upper limit evaluation. \\

The incorporation of all systematic uncertainties increases the upper 
limit by 2.1\%.
As a result, the upper limit on the branching fraction is 
\begin{equation}
{\cal{B}}(\TEG) < 3.9 \times 10^{-7} ~~~~~~ {\rm{at~90\%~C.L.}}
\end{equation}
\section{Summary}
This result improves the sensitivity to the branching fraction by 
approximately one order of magnitude compared to 
previous measurements. 
Despite a smaller detection efficiency compared to $\TMG$,
the superior BG rejection for electrons allows us
to reach a sensitivity for $\TEG$ that is comparable
to $\TMG$.

\section*{Acknowledgements}
We thank the KEKB group for the excellent operation of the
accelerator, the KEK cryogenics group for the efficient
operation of the solenoid, and the KEK computer group and
the National Institute of Informatics for valuable computing
and Super-SINET network support. 
We are grateful to A.Ilakovac for fruitful discussions.
We acknowledge support from
the Ministry of Education, Culture, Sports, Science, and
Technology of Japan and the Japan Society for the Promotion
of Science; the Australian Research Council and the
Australian Department of Education, Science and Training;
the National Science Foundation of China under contract
No.~10175071; the Department of Science and Technology of
India; the BK21 program of the Ministry of Education of
Korea and the CHEP SRC program of the Korea Science and
Engineering Foundation; the Polish State Committee for
Scientific Research under contract No.~2P03B 01324; the
Ministry of Science and Technology of the Russian
Federation; the Ministry of Education, Science and Sport of
the Republic of Slovenia;  the Swiss National Science Foundation; the National Science Council and
the Ministry of Education of Taiwan; and the U.S.\
Department of Energy.

\end{document}